\documentclass[smallextended]{svjour3}  
\usepackage[hidelinks]{hyperref}
\usepackage{arydshln}
\usepackage{color}
\usepackage{tcolorbox}
\usepackage{paralist}
\usepackage{multirow}
\usepackage{xspace}
\usepackage{algorithm}
\usepackage{algorithmic}
\usepackage{booktabs}

\usepackage{caption}
\usepackage{subcaption}

\newcommand{\newc}[1]{{\textcolor{blue}{#1}}}

\newcommand{\approach}{{\sc xTestCluster}\xspace}

\newcommand{\notpurecluster}{not pure\xspace}
\newcommand{\patchreviewer}{code reviewer\xspace}
\newcommand{\bugsmultipleclusters}{78\xspace}

\newcommand\version{1.5.0}

\begin{document}

\title{Test-based Patch Clustering for Automatically-Generated Patches Assessment}

\author{Matias Martinez   \and  Maria Kechagia   \and Anjana Perera  \and Justyna Petke \and Federica Sarro \and Aldeida Aleti}

\institute{Matias Martinez  \at Universitat Politècnica de Catalunya, Spain\\
              \email{matias.martinez@upc.edu}           
  \and
          Maria Kechagia \at University College London, United Kingdom\\
             \email{m.kechagia@ucl.ac.uk}      
              \and
          Anjana Perera \at Monash University, Australia\\
             \email{Anjana.Perera@monash.edu}     
                  \and
          Justyna Petke \at University College London, United Kingdom\\
             \email{j.petke@ucl.ac.uk}   
                        \and
                Federica Sarro \at University College London, United Kingdom\\
             \email{f.sarro@ucl.ac.uk}
                   \and
          Aldeida Aleti \at Monash University, Australia\\
             \email{aldeida.aleti@monash.edu}   
}

\maketitle

\begin{abstract}

Previous studies have shown that Automated Program Repair ({\sc apr}) techniques suffer from the overfitting problem. Overfitting happens when a patch is run and the test suite does not reveal any error, but the patch actually does not fix the underlying bug or it introduces a new defect that is not covered by the test suite. Therefore, the patches generated by {\sc apr} tools need to be validated by human programmers, which can be very costly, and prevents {\sc apr} tool adoption in practice.
%
Our work aims to minimize the number of plausible patches that programmers have to review, thereby reducing the time required to find a correct patch.
%
%
%
We introduce a novel light-weight test-based patch clustering approach called \approach, which clusters patches based on their dynamic behavior. 
\approach is applied after the patch generation phase in order to analyze the generated patches from one or more repair tools and to provide more information about those patches for facilitating patch assessment.
The novelty of \approach lies in using information from execution of newly generated test cases to cluster patches generated by multiple APR approaches.
A cluster is formed of patches that fail on the same generated test cases.
The output from \approach gives developers 
\begin{inparaenum}[\it a)]
\item a way of reducing the number of patches to analyze, as they can focus on analyzing a sample of patches from each cluster,
\item additional information (new test cases and their results) attached to each patch.
\end{inparaenum}
After analyzing 902 plausible patches from 21 Java {\sc apr} tools, our results show that \approach is able to reduce the number of patches to review and analyze with a median of 50\%.
%
\approach can save a significant amount of time for developers that have to review the multitude of patches generated by {\sc apr} tools, and provides them with new test cases that expose the differences in behavior between generated patches.
Moreover, \approach can complement other patch assessment techniques that help detect patch misclassifications.

\end{abstract}

\section{Introduction}\label{sec:introduction}

Automated program repair ({\sc apr}) techniques generate patches for fixing software bugs automatically~\cite{gazzola2017automatic,monperrus2018automatic}.
The aim of {\sc apr} is to significantly reduce the manual effort required by developers to fix software bugs. However, it has been shown that {\sc apr} techniques tend to produce more incorrect patches than correct ones~\cite{le2018overfitting,long2016analysis,martinez2017automatic}.
This issue is also known as the {\it overfitting} (or test-suite-overfitting) problem.
Overfitting happens when a patch generated automatically passes all the existing test cases  yet it fails in presence of other inputs which are not captured by the given test suite~\cite{smith2015cure}.
This happens because the test cases, which are used as program specification to check whether the generated patches fix the bug may be insufficient to fully specify the correct behavior of a program.
As a result, a generated patch may pass all the existing tests (i.e., the patch can be a {\it plausible} fix), but still be incorrect~\cite{qi2015analysis}.

Due to the overfitting problem, developers have to \emph{manually assess} the generated patches before integrating them to the code base.
Manual patch assessment is a very time-consuming and labor-intensive task~\cite{Ye2021DDR}, especially when multiple plausible patches are generated for a given bug~\cite{le2018overfitting,martinez2018ultra}.
To alleviate this problem, different techniques for {\it automated patch assessment} have been proposed. Filtering of overfitted patches can happen during patch generation, as part of the repair process (e.g.~\cite{prophet}), or as part of the post-processing of the generated patches (e.g.~\cite{ods,patchsim}).
Typically, such techniques focus on the prioritization of patches.
The patches ranked at the top are deemed to be the most likely to be correct. Existing approaches often rank similar patches at the top~\cite{le2018overfitting} and as a result waste developers' time if the top-ranked patches are overfitted.
Furthermore, such approaches might require an oracle~\cite{xin2017identifying}, or (often expensive) program instrumentation~\cite{patchsim}, or a more sophisticated machine learning process~\cite{ods}.

To alleviate these issues, we present a light-weight patch post-processing technique, named \approach, that aims to reduce the number of generated patches that a developer has to assess.
Our technique clusters plausible repair patches exhibiting the same behavior (according to a given set of test suites), and provides the developer with fewer patches, each representative of a given cluster, thus ensuring that those patches exhibit different behavior.
Our technique can be used not only when a single tool generates multiple plausible patches for a given bug, but also when different available {\sc apr} tools are running (potentially in parallel) in order to increase the chance of finding a correct patch. In this way, developers will only need to examine one patch, representative of a given cluster, rather than all, possibly hundreds, of patches produced by {\sc apr} tools. 

Our approach presents two main novelties: 
First, it leverages the diversity of the behavior of the generated patches (and this diversity is not exposed by the developer-written test cases used to synthesize patches).
In particular, our clustering approach \approach  exploits automatically generated test cases that enforce diverse behavior in addition to the existing test suite, as opposed to previous work (including Mechtaev et al.~\cite{Mechtaev:test-equivalence} for patch generation and Cashin et al.~\cite{Cashin2019Undestanding} for patch assessment) that use solely the existing test suite written by developers. 
Second, our approach has the main advantage that it does not involve code instrumentation (aside from patch application) nor an oracle (e.g., \cite{xin2017identifying}) or pre-existing dataset to learn fix patterns (e.g., \cite{ods,Lin2022ContextEmbeeding}).  
Moreover, \approach is complementary to previous work on patch overfitting assessment, as it can apply different prioritization strategies to each cluster.

\approach works as follows: 
First, \approach receives as input a set of plausible patches generated by a number of selected {\sc apr} tools and it generates new test cases for the patched files (i.e., buggy programs to which plausible patches have been applied) using automated test-case generation tools. 
The goal of this step is to generate new inputs and assertions that expose the behavior (correct and/or incorrect) of each generated patch.
Second, \approach executes the generated test cases on each patched file to detect any behavioral differences among the generated patches (we call this step  \textit{cross test-case execution}).
Third, \approach receives the results from the execution of each test case on a patched version of a given buggy program, and uses the names of the failing test cases to cluster patches together.
In other words, patches from the same cluster exhibit the same output, according to the generated test cases: they fail on all the same generated tests. Patches that pass all the test cases (no failing tests) are clustered together.

We evaluate our approach on 902 patches (248 correct and 654 overfitted) for bugs from v.\version~of the {\sc Defects4J} data set~\cite{Just2014Defects4J}, generated by 21 different {\sc apr} tools, and collected and labeled by Wang et al.~\cite{wang2020AutomatedAssessment}. 
After removing duplicates, we used two automated test-case generation tools, {\sc EvoSuite}~\cite{evosuite} and {\sc Randoop}~\cite{randoop}, to generate test cases for our patch set.
Finally, we cluster patches based on test case results.
To our knowledge, \approach is the first approach to analyze together patches from multiple program repair approaches generated to fix a particular bug.
This is important because it shows that \approach can be used in the wild, independently of the adopted Java repair tools.

Our results show that \approach is able to create at least two clusters for almost half of the bugs that have two or more different patches.
By having patches clustered, \approach is able to reduce a median of 50\% of the number of patches to review and analyze. 
This reduction could help code reviewers (developers using automated repair tools or researchers evaluating patches) to reduce the time of patch evaluation.
Additionally, \approach can also provide code reviewers with the inputs (from the generated test cases) that trigger different program behaviors for different patches generated for one bug.
This additional information may help them decide which patch to select and merge into their codebase.

We also analyze the assessment done by two state-of-the-art patch assessment approaches, ODS~\cite{ods} and Cache~\cite{Lin2022ContextEmbeeding} on the patches clustered by \approach.
The results show that \approach can be used complementarily to those approaches and can help to detect false positives and false negatives.
Finally, we study whether metrics related to the quality of the newly-generated tests in order are related to the efficiency of \approach. 
We found that the number of test cases generated, their lengths (in terms of lines of code), and the coverage affect the ability to cluster correct patches together.

Overall, the paper provides the following contributions:
\begin{itemize}
    \item A novel test-based patch clustering approach called \approach.
It is complementary to existing patch overfitting assessment approaches.
\approach can be applied to patches generated by multiple APR tools.
    \item An implementation of \approach for analyzing Java patches. 
    It uses two popular automated test-case generation frameworks, {\sc EvoSuite}~\cite{evosuite} and {\sc Randoop}~\cite{randoop}. 
    The code of \approach is publicly available~\cite{appendix}.
    \item An evaluation of \approach using patches from 21 {\sc apr} tools, and 920 plausible patches.
\end{itemize}

All our data is available in our appendix~\cite{appendix}.

\section{Our approach}
\label{sec:algorithm}

\begin{figure}
\centering
\includegraphics[width=0.95\textwidth]{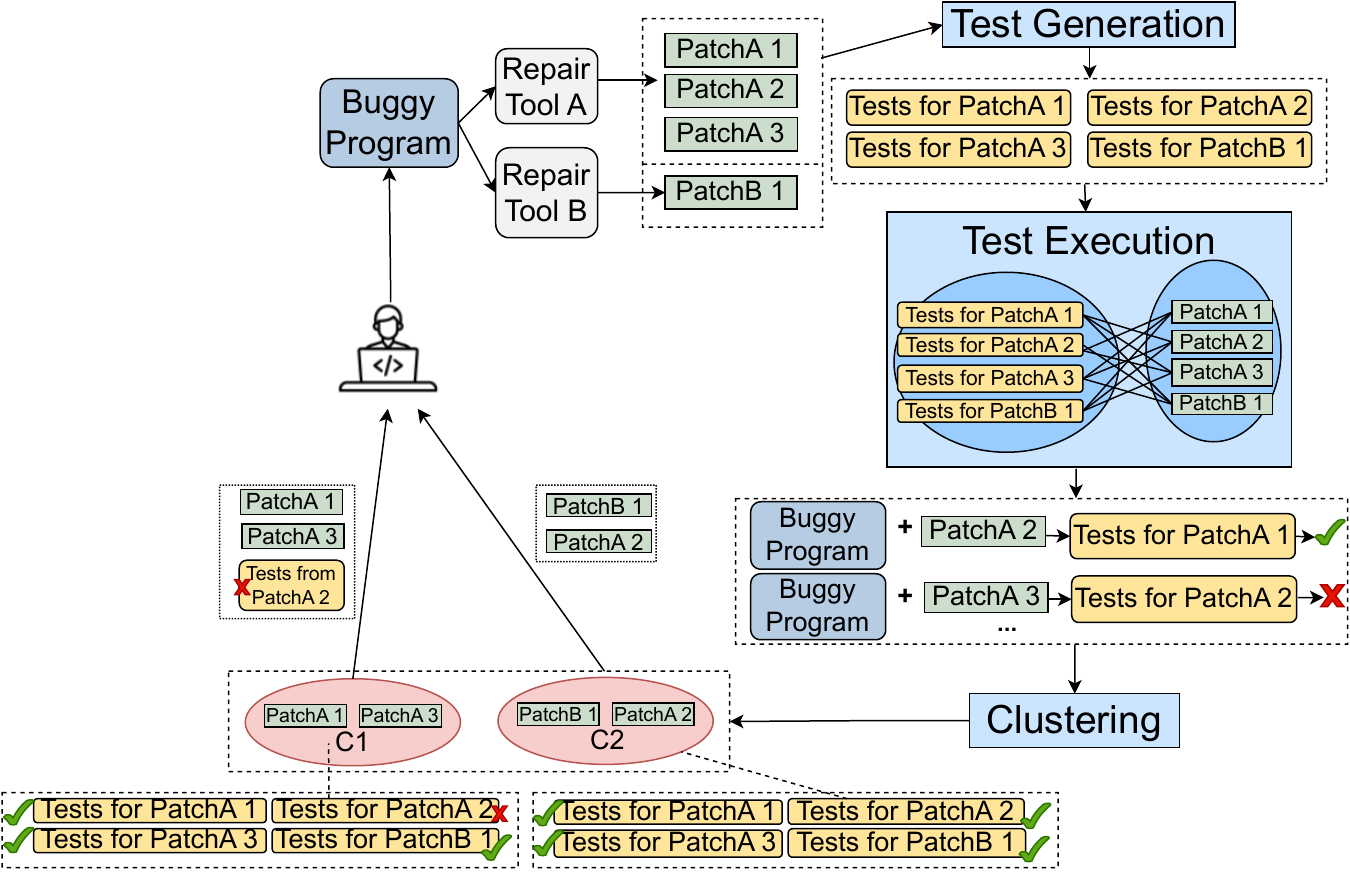}
 \centering
    \caption{The steps executed by \approach.}
    \label{fig:approach}
\end{figure}

Our proposed approach, \approach,
for test-based patch clustering is shown in Figure~\ref{fig:approach}.


 \begin{algorithm}[!ht]
   \caption{\approach}
   \begin{algorithmic}[1]
   \STATE \textbf{Input:}  
    $B$ a buggy program,
    $Ps$ plausible patches of bug $B$,
    $TGs$ test-case generators.
    
    \STATE $TCG \gets testGeneration(B, Ps, TGs)$ \COMMENT{(Alg. \ref{algo:testGeneration})}
   \STATE $ResPatches_{exec} \gets testExecution(B, Ps, TCG)$  \COMMENT{(Alg. \ref{algo:testExecution})}
    \STATE $clusters, exec\_info \gets clustering(Ps, ResPatches_{exec})$  \COMMENT{(Alg. \ref{algo:clustering})}
   \RETURN $clusters, exec\_info$ 
   \label{algo1:general:return}
   \end{algorithmic}
   \label{algo:CompleteAlgo}
 \end{algorithm}


\approach receives as input a buggy program and a set of plausible patches that could repair the bug. 
The patches could have been automatically generated by one or multiple repair approaches.
Additionally, \approach receives a set of test-case generation tools.
Given these inputs, \approach carries out three steps (lines 2--5 of Algorithm~\ref{algo:CompleteAlgo}):
\begin{inparaenum}[\it 1)]
\item test-case generation,
\item test-case execution, and
\item clustering. 
\end{inparaenum}

{\bf Test-case generation.} 
\approach receives as input the plausible patches generated by a set of {\sc apr} tools and generates new test cases for the patched files (i.e., buggy programs to which plausible patches were applied to).
We use automated test case generation tools for this purpose.

{\bf Test-case execution.} 
\approach executes the generated test cases on each patched version of the program: We call this approach {\it cross test-case execution}, because the test cases generated for a given patched program version are executed on another patched version of a given buggy program.
This cross execution aims to detect the behavioral differences among the generated patches that we use in clustering afterward.

{\bf Clustering.} \approach receives the results from the execution of each test case on a patched version of a given buggy program and uses this information to cluster patches together: if two patches have the same output for all the test cases generated, then they belong to the same cluster.
Each cluster of patches will also be furnished with test cases for which the patches fail, and the corresponding failures observed.

These steps allow \approach to reduce the number of patches that are presented to the {\it \patchreviewer}.
A \patchreviewer could be, for example, a software developer that has developed and pushed a buggy version, which is exposed, for example, through failed test cases executed by a continuous integration platform ({\sc ci}). 
Without using \approach, the \patchreviewer needs to assess patches produced by repair approaches that they have integrated, for example, in their {\sc ci}.
Using \approach, the \patchreviewer can, now, review a {\it subset} from all the generated patches, reducing the review effort.
Moreover, for each presented patch, they also have alternative patches (those not selected from the same cluster but with the same behavior as the selected patch) and information about test-case executions. 
All this information could help \patchreviewer decide which patch to integrate into the codebase to fix the given bug.

In the following subsections, we detail each step of \approach.

\subsection{Test-case Generation}
\label{sec:approach:testgeneration}
 \begin{algorithm}[t]
   \caption{Test-case Generation}
   \begin{algorithmic}[1]
   \STATE \textbf{Input:}  
    $B$ a buggy program,
    $Ps$ plausible patches of bug $B$,
    $TGs$ test-case generators.
   
   \STATE $TCG \gets \emptyset$ 
   \FOR{$patch \in Ps $} \label{algo1:tg:for:patch}
    \STATE $B' \gets$  apply $patch$ to $B$ \label{algo1:tg:applypatch}
    \STATE $pfiles \gets getFiles(patch)$\label{algo1:tg:getfilespatched}
     \FOR{$tg \in TGs $} \label{algo1:tg:for:testgen}
     \FOR{$pfile \in pfiles $} \label{algo1:tg:for:file}
       
         \STATE $TC_{new} \gets generateTests(tg, B', pfile)$   \label{algo1:tg:generate}
         \STATE $TCG \gets TCG \cup TC_{new} $ \label{algo1:tg:addtests}
         
      \ENDFOR  
     \ENDFOR  
    \ENDFOR  
   
   \RETURN $TCG$ \label{algo1:tg:return}
   \end{algorithmic}
   \label{algo:testGeneration}
 \end{algorithm}

Algorithm \ref{algo:testGeneration} details this step.
For each patch  $patch$ from those plausible patches received as input (line \ref{algo1:tg:for:patch}), 
\approach first applies this patch
to the buggy program $B$, giving the patched program $B'$ as a result (line \ref{algo1:tg:applypatch}).
We recall that the patched program must pass all test cases provided by the developer.
If the patch does not pass any of those test cases, it is not plausible, and \approach discards it. 
Then, \approach retrieves the files affected by the patch (line \ref{algo1:tg:getfilespatched}). 
Using each of the test-case generation tools (line \ref{algo1:tg:for:testgen}) we have selected, 
\approach generates test cases for each of those files that have plausible fixes (line \ref{algo1:tg:generate}).
All the generated test cases are stored in a set called $TCG$ (line \ref{algo1:tg:addtests}).
\approach is agnostic to the test case generation tools employed in practice. This means that, in theory, any test case generation tool could be used.
For this reason, in this section we do not detail the implementation of the $generateTest$ invocation at line \ref{algo1:tg:generate}, i.e., how tests are generated.
Nevertheless, in the Methodology section (\ref{sec:methodology:rq1clusters}) we detail implementation of our approach used in the evaluation (which is based on two state-of-the-art test generation tools: Evosuite~\cite{evosuite} and Randoop~\cite{randoop}).

\approach also carries out a {\it sanity check} on the generated test cases. 
In particular, it verifies that they are not flaky by executing them $n$ times, and assuring that the results are the same for each execution. 
Test cases that do not pass this check are discarded.\footnote{For simplicity we do not show this check in Algorithm \ref{algo:testGeneration}.}

\subsection{Test-case Execution}
\label{sec:approach:testexecution}
 \begin{algorithm}[t]
   \caption{Test-case Execution}
   \begin{algorithmic}[1]
   \STATE \textbf{Input:}  
   $B$ a buggy program ,
    $Ps$ plausible patches of bug $B$,
    $TCG$ test cases generated.
    
     
    \STATE $ResPatches_{exec} \gets \emptyset $
    \FOR{$patch \in Ps $} \label{algo1:te:for:patch}
        \STATE $R_{exec} \gets \emptyset $
        \STATE $B' \gets$  apply $patch$ to $B$ \label{algo1:te:applypatch}
        \FOR{$t \in TCG $} \label{algo1:te:for:testgenerated}
         \STATE $res_t \gets execute(t, B', patch)$ \label{algo1:te:testexec}
          \STATE $R_{exec} \gets R_{exec} \cup (res_t, patch)$ 
         \ENDFOR  
         \STATE $ResPatches_{exec} \gets ResPatches_{exec}  \cup \langle patch, R_{exec}\rangle $  \label{algo1:te:storeresult}
    \ENDFOR  

   \RETURN $ResPatches_{exec} $ \label{algo1:te:result}
   \end{algorithmic}
   \label{algo:testExecution}
 \end{algorithm}

Next, we conduct \textit{cross test-case execution}. 
Algorithm \ref{algo:testExecution} details this step.
\approach executes, on a version of the program patched with the patch $patch$, the test cases generated by considering other plausible patches for bug $B$.
In other words, the step applies the Cartesian product between patches and test cases produced for the patches.
To achieve this, \approach iterates over the patches (line \ref{algo1:te:for:patch}).
For each patch, \approach applies it to the buggy program, producing the patched program $B'$ as a result (line \ref{algo1:te:applypatch}).
\approach executes each new test case $t$ generated in the previous step (set from $TCG$) over the patched program $B'$ (lines \ref{algo1:te:for:testgenerated} and \ref{algo1:te:testexec}).
All results from the test-case execution for $patch$ are stored in the map $ResPatches_{exec}$ (line \ref{algo1:te:storeresult}), which is then returned (line \ref{algo1:te:result}).

\subsection{Clustering}
\label{sec:approach:clustering}
 \begin{algorithm}[t]
   \caption{Clustering}
   \begin{algorithmic}[1]
   \STATE \textbf{Input:}  
    $Ps$ plausible patches of bug $B$,
    $ResPatches_{exec}$  results of test cases.
    
    \STATE $Cs \gets \emptyset$ 
    \STATE $FailsCs \gets \emptyset$ 
    
    \FOR{$patch_i \in Ps $} \label{algo1:clust:for:patch}
        \IF{$Cs$ is $\emptyset$ } 
        \label{algo1:clust:nocluster}
            \STATE $Cs \gets set(patch_i)$  \label{algo1:clust:initcluster}
            \STATE $FailsCs \gets getTestExecution(ResPatches_{exec}, patch_i)$ 
            \label{algo1:clust:initfailures}   
        \ELSE
            \STATE $R_{exec_i} \gets getTestExecution(ResPatches_{exec}, patch_i)$  \label{algo1:clust:getexec}
            \STATE  $ foundCluster  \gets false$ 
               \FOR{$cluster \in C_s$} \label{algo1:clust:searchcluster}\label{algo1:clust:for:cluster}
                   \STATE $ patch_o \gets getOne(cluster) $  \label{algo1:clust:getonepatch}
                    \STATE $R_{exec_o} \gets getTestExecution(ResPatches_{exec}, patch_o) $  \label{algo1:clust:getexecother}
                   
                    \IF{$R_{exec_i} = R_{exec_o} $ } \label{algo1:clust:foundcluster}
                         \STATE $cluster \gets cluster \cup patch_i$ 
                         \label{algo1:clust:addpatchtocluster}
                           \STATE  $ foundCluster  \gets true $
                            \STATE break       \label{algo1:clust:break}
                    \ENDIF    
                \ENDFOR 
                \IF{foundCluster = false} \label{algo1:clust:ifnotfoundcluster}
                  \STATE $Cs \gets Cs \cup set(patch_i)$ \label{algo1:clust:notfoundcluster:create}
                   \STATE $FailsCs \gets FailsCs \cup R_{exec_i}$ \label{algo1:clust:notfoundcluster:addfailure}
                \ENDIF
         \ENDIF

    \ENDFOR  
   
   \RETURN $Cs$, $FailsCs$ \label{algo1:clust:returnclusters}
   \end{algorithmic}
   \label{algo:clustering}
 \end{algorithm}

Algorithm \ref{algo:clustering} details this step.
\approach iterates over the patches (line \ref{algo1:clust:for:patch}) in order to assign each patch $patch_i$ to a cluster.
If no cluster has been previously created (line \ref{algo1:clust:nocluster}), \approach creates a new cluster that includes $patch_i$ (line \ref{algo1:clust:initcluster}), and stores the results of the test cases (i.e., the failing test cases if any) in the list $FailsCs$  (line \ref{algo1:clust:initfailures}).
The $i$-th element of that list contains the execution results of the $i$-th cluster.
Otherwise, \approach first retrieves the results from the \emph{test-case execution} step (Algorithm \ref{algo:testExecution}) for that patch (line \ref{algo1:clust:getexec}).
Then, \approach iterates over the created clusters (line \ref{algo1:clust:for:cluster}).
For each $cluster$, \approach chooses a patch $patch_o$ from it (line \ref{algo1:clust:getonepatch}) and retrieves the corresponding results from the test case execution step (line \ref{algo1:clust:getexecother}).
\approach compares the two execution results (line \ref{algo1:clust:foundcluster}):
If both patches produce the same failures in our test set after being applied to the buggy program,
the patch $patch_i$ is included in $cluster$ (line \ref{algo1:clust:addpatchtocluster}).
Note that the result of a test case can be \emph{passing} or \emph{failing}.
When the result is failing, we also consider the message associated with the failing assertion or the message associated with an error.
Consequently, patches that do not pass a test case due to different reasons (e.g., some fail an assertion and others produce a {\tt null} pointer exception) are allocated into different clusters.
Patches that pass all test cases (this means that $R_{exec_o}$ at line \ref{algo1:clust:getexecother} is empty) are clustered together.
If \approach cannot allocate $patch_i$ to any cluster (line \ref{algo1:clust:ifnotfoundcluster}), it creates a new cluster which includes $patch_i$ (line \ref{algo1:clust:notfoundcluster:create}) and stores the test execution result in $FailsCs$ (line \ref{algo1:clust:notfoundcluster:addfailure}).

Finally, \approach returns all the created clusters $Cs$ and a list with the test case execution (i.e., failing cases) for each of those clusters (line \ref{algo1:clust:returnclusters}).

\section{Research Questions}
\label{sec:researchquestions}

Our proposal, \approach, aims to aid code reviewers
in reducing the effort required for manual assessment
of patches automatically generated by {\sc apr} tooling.
In order to assess how well \approach can achieve this task we pose the following RQs:

\newcommand{\rqcluster}{RQ1: Hypothesis Validation}

\begin{quote}
\textbf{\rqcluster} \textit{To what extent are generated test cases able to capture behavioral differences between patches generated by {\sc apr} tools?}
\end{quote}

\noindent This RQ aims to show the ability of \approach to detect behavioral differences among the generated patches based on the execution of generated test cases, and, from those differences, to create clusters of patches.
If successful, semantically equivalent patches will be clustered together and only one, thus, needs to be presented to \patchreviewer from such a cluster.

\newcommand{\rqeffortreduction}{RQ2: Patch Reduction Effectiveness}
\begin{quote}
\textbf{\rqeffortreduction} \textit{How effective is \approach at reducing the number of patches that need to be manually inspected? }
\end{quote}


\noindent This RQ aims to show the applicability of \approach in order to help developers reduce the effort of manually reviewing and assessing patches.
In particular, we compare the number of patches produced by all selected APR tools vs.\ the number of patches presented to \patchreviewer if our approach is used.

\newcommand{\rqcorrectcluster}{RQ3: Clustering Effectiveness}

\begin{quote}
\textbf{\rqcorrectcluster} \textit{How effective is \approach at clustering correct patches together?}
\end{quote}

\noindent This RQ aims to measure the ability of \approach to cluster correct patches together.
In case each cluster contains {\it only} either correct 
or incorrect patches, we can simply pick {\it any} patch from a given cluster for validation.
In other words, picking {\it any} patch from a cluster would be sufficient to ensure existence (or non-existence) of a correct patch among {\it all} plausible patches in a cluster during patch assessment.


\newcommand{\rqsota}{{RQ4:} Complementing State-of-the art patch assessment techniques}
\begin{quote}
\textbf{\rqsota} \textit{To what extent is \approach able to complement existing overfitting detection techniques in order to help them to reduce the rate of incorrect assessments?}
\end{quote}

This RQ aims to study the assessment done by state-of-the-art patch assessment approaches on the clusters found by \approach.
We will specially focus on mixed clusters, i.e. those that contain both correct and incorrect patches, and we study the false positives and false negatives of other patch assessment approaches on patches assigned by \approach to those clusters.

\newcommand{\rqqualitytc}{RQ5: Quality of generated test cases and Effectiveness of \approach}
\begin{quote}
\textbf{\rqqualitytc} \textit{
To what extent does quality of the newly-generated test cases affect \approach's effectiveness?
}
\end{quote}

This RQ aims to study the relationship between the quality of test cases (represented using metrics extracted from the newly generated test cases) and the effectiveness of \approach to differentiate correct patches from incorrect.
In particular, we want to know if the limitations of \approach are due to the quality of the generated tests it uses for comparing the behaviours.

\section{Methodology}
\label{sec:methods}

In this section, we present the methodology followed to answer our research questions.

\subsection{Dataset}
\label{sec:metodology:datasets}

In order to evaluate \approach we need a set of plausible patches, i.e., proposed fixes for a given bug.
There are two constraints the dataset needs to meet.
Firstly, as \approach focuses on the reduction of the amount of patches to be presented to the developers for review, \approach makes sense only if there are at least two different plausible patches for a given bug.
Secondly, we need to know whether each patch in the dataset is correct or not.
In previous work (e.g., \cite{Ye2021DDR}) patches have been labelled (usually via manual analysis) as either \emph{correct}, \emph{incorrect} (or \emph{overfitting}), or marked as \emph{unknown} (or \emph{unassessed}). 
We will use the \emph{correct} and \emph{incorrect} label terminology, and only consider patches for which correctness has been established.

We thus consider patches generated by existing repairs tools. 
In this experiment, we focus on tools that repair bugs in Java source code because:
\begin{inparaenum}
\item Java is a popular programming language, which we aim to study, and 
\item the most recent repair tools have been evaluated on bugs from Java software projects.
\end{inparaenum}

Since the execution of {\sc apr} tools to generate patches is very time-consuming (especially if we consider several repair tools~\cite{Durieux:2019:RepairThemAll}), we use publicly available patches that were generated in previous APR work.
We also decide to rely on external patch evaluations done by other researchers and published as artifacts to peer-reviewed publications.
This avoids possible researcher bias. 
Furthermore, it allows us to gather a large dataset of patches, from multiple sources, and avoid the costly manual effort of manual patch evaluation of thousands of patches.

Taking all constraints into account, we decided to study  patches for bugs from the {\sc Defects4J}~\cite{Just2014Defects4J} dataset. 
To the best of our knowledge, {\sc Defects4J} is the most widely used dataset for the evaluation of repair approaches~\cite{martinez2017automatic,Durieux:2019:RepairThemAll,Liu2020OnTE}. 
Consequently, hundreds of patches for fixing bugs in {\sc Defects4J} are publicly available.
We leverage data from previous work that has collected and aggregated patches generated by different Java repair tools, all evaluated on {\sc Defects4J}.
We focus on those that, in addition, provide a {\it correctness label}.

\begin{table}[]
\centering
\begin{tabular}{lrr}
\toprule
\multirow{3}{*}{{\bf Tools}} &\multicolumn{2}{c}{{ Dataset of Patches}}\\
\cline{2-3}
&\multicolumn{2}{c}{Wang et al.~\cite{wang2020AutomatedAssessment}}     \\ 
& {Correct} & Overfitted \\ 
\midrule
ACS~\cite{acs2017} & 15  &  7 \\
AVATAR~\cite{LKK19} & 19  &  38 \\
Arja~\cite{YB18} & 5  &  52 \\
CapGen~\cite{capgen} & 25  &  41 \\
Cardumen~\cite{MM16} & 2  &  10 \\
DynaMoth~\cite{DM16} & 1  &  21 \\
FixMiner~\cite{KLB20} & 12  &  20 \\
GenProg-A~\cite{YB18} & 2  &  28 \\
Jaid~\cite{CPF17} & 40  &  41 \\
jGenProg~\cite{MM16} & 3  &  17 \\
jKali~\cite{MM16} & 3  &  22 \\
jMutRepair~\cite{MM16} & 5  &  17 \\
Kali-A~\cite{YB18} & 3  &  60 \\
kPAR~\cite{kpar2019} & 10  &  52 \\
Nopol~\cite{XMD17} & 1  &  30 \\
RSRepair-A~\cite{YB18} & 2  &  39 \\
SOFix~\cite{LZ18} & 21  &  2 \\
SequenceR~\cite{CKT19} & 17  &  56 \\
SimFix~\cite{JXZ18} & 22  &  46 \\
SketchFix~\cite{HZW18} & 16  &  7 \\
TBar~\cite{LKK19b} & 24  &  48 \\
\hline
Total & 248  &  654 \\
\bottomrule
\end{tabular}
\caption{Number of Correct and Overfitted patches for bugs from {\sc Defects4J}~\cite{Just2014Defects4J} per repair tool, contained in the dataset from Wang et.~\cite{wang2020AutomatedAssessment} composed of 902 patches.}
\label{tab:summaryTools}

\end{table}

In this paper, we use the dataset provided by Wang et al.~\cite{wang2020AutomatedAssessment} which contains 902 patches from 21 repair systems.
We choose this dataset for the following main reasons.
\begin{itemize}
    \item Firstly, it met all our criteria mentioned above (i.e., Java patches,  Defects4J patches, correctness labels).
   \item  Secondly, it is a combination of other two datasets of labeled patches, one from Liu et al.~\cite{Liu2020OnTE} and the second from {\sc DDR} by He~et~al.~\cite{Ye2021DDR}. 
Liu et al.~\cite{Liu2020OnTE} included patches from 16 repair systems, and manually evaluated the correctness using guidelines presented by Liu et al.~\cite{Liu2020OnTE}.
He~et~al.~\cite{Ye2021DDR} classified patches using a technique called {\sc RGT}, which generates new test cases using ground-truth, human-written oracle patches.
The patches from these datasets were revised by Wang and colleagues in order to correct existing correctness label misclassifications. 
\item Lastly, the dataset from Wang et al. has been widely used in the evaluation of overfitting approaches, including~\cite{ods,Lin2022ContextEmbeeding,wang2020AutomatedAssessment}. 
Using this dataset allows us to compare the performance of \approach with previous work.
\end{itemize}
Table \ref{tab:summaryTools} presents the number of patches from the Wang et al. dataset per repair tool. The dataset contains 902 bugs, 248 were labeled as correct, and the other 654 as overfitting (incorrect).

\begin{table}[]

\centering
\begin{tabular}{lr}
\toprule
{\bf Summary of Bugs} & {\bf \#Bugs}\\
\midrule
Total bugs from {\sc Defects4J}~\cite{Just2014Defects4J}  & 375 \\ 
Total bugs with labelled patches  & {202} \\ 
Bugs with one patch       & {70} \\ 
Bugs with $>$ distinct one patch      & {132}\\ 
Bugs considered by \approach    & {129}   \\
\bottomrule
\end{tabular}
\caption{Summary of bugs from {\sc Defects4J} and their respective patches collected from the three datasets.}
\label{tab:summaryBugs}
\end{table}

We now explain how we processed the Wang et al. datasets in order to select the bugs and patches that we are interested in analyzing.
Table \ref{tab:summaryBugs} presents the number of bugs for which patches were generated.
In total, 202 bugs from the version \version~of {\sc Defects4J} contain at least one patch on  Wang et al.'s dataset.

For each bug, we carry out a syntactic analysis of patches (using the \emph{diff} command) in order to detect duplicate patches produced by two or more repair tools. 
This is necessary, as multiple tools can create exactly the same patch. 
In total, we consider 777 distinct patches.

From the patches gathered for 202 bugs in our dataset, we find that for 70 bugs there is only one single patch.
We discard those bugs and their patches because \approach needs at least two patches per bug.
We also discard patches for three further bugs for the following reasons:
First, bugs Closure-63 and Closure-93 were deprecated in a recent version of Defects4J.\footnote{\url{https://github.com/rjust/defects4j\#the-projects}}
Consequently, the Defects4J framework does not allow us to checkout and test those bugs.
Second, for Math-35 bug, we could generate tests for only one patch, so we discarded it.

In total, we evaluate \approach on 129 bugs, each having at least two patches which we can successfully generate test cases for.

\subsection{\rqcluster}
\label{sec:methodology:rq1clusters}

In order to answer RQ1, we group patches by bug repaired by the tools.
Then, we apply the algorithm described in Section \ref{sec:approach:testgeneration}.
In this experiment, we report the results obtained by using Evosuite~\cite{evosuite} as the test-case generation tool.
The results by adding test cases from Randoop~\cite{randoop} are discussed in Section~\ref{sec:discussion:testgeneration}.
For each bug from {\sc Defects4J}, we generate test cases for each patched version, i.e., after applying a candidate patch to the buggy program, using both tools and a time budget of 60 seconds for the test-case generation.\footnote{In the Evosuite code, the default search budget is set to 60 seconds (\url{https://github.com/EvoSuite/evosuite/blob/1948d763944b3c3275d9564cf375b31981aedab0/client/src/main/java/org/evosuite/Properties.java\#L690}), however we found an official documentation mentioning a different value (10 minutes, \url{https://www.evosuite.org/documentation/commandline/)}. 
To avoid any confusion, we call Evosuite by explicitly passing the search budget of 60 seconds.
}
As this approach aims to not depend on any oracle (inc. human-written test cases), we trigger the test generation from scratch, without providing any test cases as seed.
When invoking these generation tools, \approach uses the default values for each hyperparameter\footnote{Default values of Evosuite parameters \url{https://github.com/EvoSuite/evosuite/blob/1948d763944b3c3275d9564cf375b31981aedab0/client/src/main/java/org/evosuite/Properties.java\#L516}}.
In particular, in Evosuite, the default objective of the search is to maximize the line coverage of the test cases under generation.
Additionally, by default, EvoSuite applies minimization to generated test cases, which means that it removes all statements that are not strictly needed to satisfy the coverage goals.

Then, we execute the test cases generated on the patched versions (cross test-case execution).
Finally, we cluster patches for a single bug by putting together all the patches that have the same results on the generated test cases, as explained in Section~\ref{sec:approach:clustering}.

\subsection{\rqeffortreduction}
\label{sec:methodology:rq2reduction}

To answer RQ2 we take as input the number of clusters generated in RQ1 and the total number of patches in our dataset per bug.
We recall that these patches from a bug could come from:
\begin{inparaenum}[\it 1)]
    \item a single repair tool (which does not stop the search after the first test-passing patch is found and finds multiple such patches),
    \item multiple repair tools (where each could potentially stop after the first test-passing patch is found).
\end{inparaenum}

For each bug, we compute the reduction of patches to analyze per bug $B$ as follows:
\begin{equation}
reduction(B) =  \frac{(\#patches \; for \; B -  \#clusters \; for  \; B )}{ \#patches \; for \; B} \times 100
\end{equation}

This gives us the \% reduction of patches presented to a \patchreviewer.
Recall that we only present one patch from each cluster to \patchreviewer, i.e, $\#clusters$ patches. 
Otherwise, \patchreviewer would have had to review $\#patches$ per bug.

\subsection{\rqcorrectcluster}
\label{sec:methodology:rq3correctness}
\label{sec:methodology:correctness}

To answer RQ3, we take as input:
\begin{inparaenum}[\it 1)]
\item the clusters generated by \approach (we recall a cluster has one or more patches), and
\item the correctness labels for each patch in our dataset (see Section~\ref{sec:metodology:datasets}).  
\end{inparaenum}
We say that a cluster is \emph{pure} if all its patches have the same label,
i.e, all patches are correct or all patches are incorrect.
Otherwise, we say the cluster is \emph{\notpurecluster}.

For each of the sets of patches per bug, we compute the ability of \approach to generate only pure clusters.
Having bugs with only pure clusters is the main goal of \approach:
If all patches in a cluster are correct, by picking one of them we are sure to present a correct one to the \patchreviewer.
Similarly, if all patches from a cluster are incorrect, by picking one of them we are sure to present to the \patchreviewer an incorrect patch.
In both cases, the reduction of patches presents no risk and patches can be picked from a cluster in any order, e.g., at random.

\subsection{\rqsota}

To answer RQ4, we study the performance of patch assessment approaches in patches clustered by \approach.
For a fair comparison, we consider approaches also evaluated on the same data used in this paper (Wang et al.~\cite{wang2020AutomatedAssessment}).
The recent empirical study by Lin et al.~\cite{Lin2022ContextEmbeeding} compared 15 patch assessment approaches (including PatchSim~\cite{patchsim}, S3~\cite{s3}, CapGen~\cite{capgen}, ssFix~\cite{ssFix}, ODS~\cite{ods} and Cache~\cite{Lin2022ContextEmbeeding}) on the mentioned dataset.
They show that ODS and Cache are the approaches with the highest \emph{accuracy}, \emph{recall} and \emph{f1} metrics. 
As their values of \emph{f1} are extremely similar, we decided to include both of them in this experiment.

We first observe their results on patches assigned to pure clusters.
That will help us identify the cases for which \approach works as expected but any of the state-of-the-art approach fails.
Then, we study their performance on patches assigned to mixed clusters. That is, on the cases for which \approach fails. 
We study whether state-of-the-art approaches also fail in such cases.
As both ODS and Cache have been evaluated on the Wang et al. dataset, composed of 902 patches, we consider the performance of these tools available in the Appendix of ODS\footnote{ODS appendix: \url{https://github.com/SophieHYe/ODSExperiment}} and Cache\footnote{Cache appendix: \url{https://github.com/Ringbo/Cache}}.

\subsection{\rqqualitytc}

To answer RQ5, we consider four metrics associated with the newly generated test cases. 
Those metrics are: 
\begin{inparaenum}[a)]
\item  number of test cases generated per patch,
\item lines of code corresponding to these test cases,
\item line coverage that reach these test cases i.e., the number of executed lines per these test cases divided by the total lines of code contained in them, and 
\item mutation score i.e., total number of mutants killed by these test cases divided by the total number of mutants evaluated.
\end{inparaenum}

To know to what extent these metrics are related to the \approach effectiveness we carry out the following steps. For each of the metrics, we first create two samples: 
The first one contains the metric values of the test cases that allow \approach to create pure clusters (in other words, those are the tests generated from patches assigned to pure clusters).
The second one contains the metric values of those test cases that are not capable of differentiating the behaviour between correct and incorrect patches (in other words, those are the tests generated from patches that belong to mixed clusters).
Then, we compare the two samples to know whether the samples have the same distribution (Null hypothesis $H_0$) or different distribution (Alternative hypothesis). 
If the Null hypothesis holds, it means that the metric does not have an impact on \approach effectiveness. 
To test the hypotheses, we use the Mann-Whitney U test, with a significance cutoff ($\alpha$ level) equal to 0.05~(5\%).


\section{Results}


\label{sec:results}

In this section we present the results of our experiments with answers to our research questions.

\subsection{\rqcluster}
\label{sec:result:cluster}

\begin{table}[]
\centering
\begin{tabular}{lr}
\toprule
{\bf Bugs under Analysis} & {\bf \#Bugs} \\ \midrule
Bugs with one or more patches                           & {202}    \\ 
Bugs with 1+ syntactically different patches                           & {129}    \\ 
Bugs with multiple patches in one single cluster      & {51} \\ 
\bf{Bugs with  multiple patches and clusters} & {78}  \\ 
\bottomrule

\end{tabular}
\caption{RQ1. Classification of bugs according to the number of patches and clusters generated by \approach.  }

\label{tab:clustersummary}
\end{table}

As Table \ref{tab:clustersummary} shows, \approach is able to generate more than one cluster for 78 bugs (60.4\%) containing more than one plausible patch (129 bugs in total).
This means that \approach, using generated test cases, is able to differentiate between patches whose application produces different behavior.

For 51 bugs (39.6\%), for which we have more than two syntactically different patches, \approach groups all of them into one cluster.
We conjecture that this could be caused by the following reasons:
\begin{inparaenum}[\it 1)]
\item Beyond syntactical differences, the patches could be semantically equivalent; 
\item test-case generation tools are not able to find inputs that expose behavioral differences between the patches;
\item test-case generation tools are not able to find the right assertion for an input that could expose behavioral differences.
\end{inparaenum}

\begin{figure}
\includegraphics[width=0.70\textwidth]{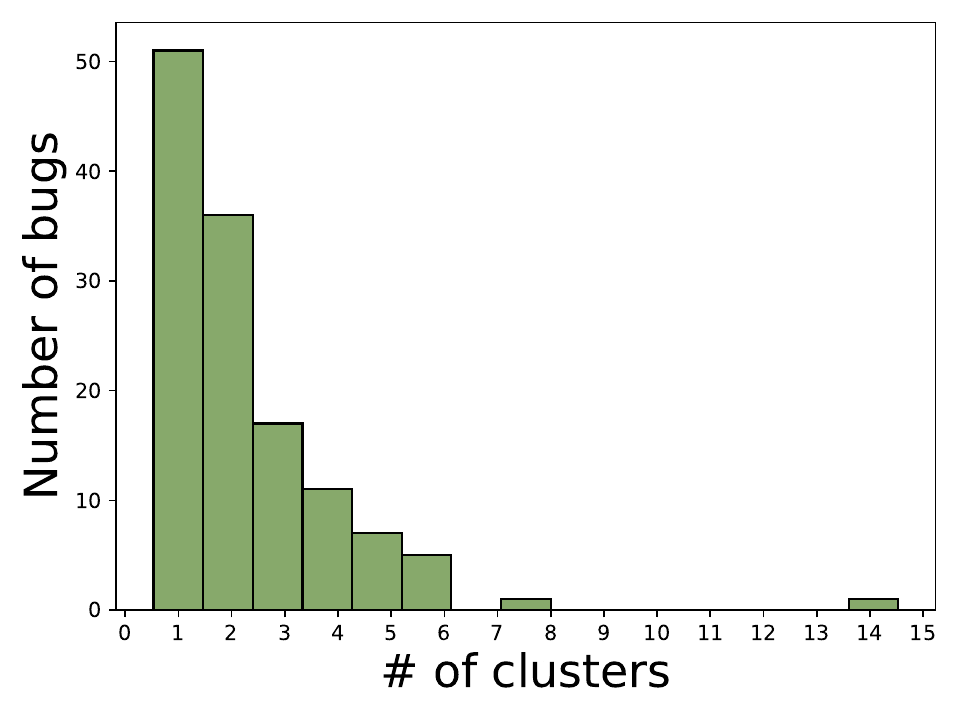}
 \centering
    \caption{RQ1. Distribution of the number of clusters per bug. Bugs with a single patch (in total 51) are discarded.}
    \label{fig:distClustersAll}
\end{figure}

Figure \ref{fig:distClustersAll} shows the distribution of the number of clusters that \approach is able to create per bug (in total, {129 bugs} as explained in Section \ref{sec:methodology:rq1clusters}).
We observe that the distribution is right-skewed.
The most left bar corresponds to the previously mentioned {51} bugs with one cluster.
Then, the number of bugs with $n$ clusters decreases as $n$ increases.
For 76 bugs, \approach generated between two and six clusters, but for 2 bugs, it generates a larger number of clusters (eight and fifteen).

\begin{tcolorbox}
  {\bf Answer to RQ1.} Given 129 bugs with at least two plausible and syntactically different patches, for {78} of them {(59.4\%)}, \approach is able to detect patches with different behavior (based on test-case generation) and group them into different clusters.
\end{tcolorbox}

By reviewing one patch per cluster, a code reviewer can reduce the time and effort required for reviewing, since he or she does not need to review all the generated patches for 59.4\% of the bugs (78 in total).
In particular, let us focus on these 78 bugs for which \approach reduces the number of patches.
Without \approach, 
the mean number of patches to be reviewed by a developer is 6.89 patches. 
On the contrary,  \approach offers a developer a median of 3.26 patches, all of which behave differently according to the generated test cases.
This means that \approach actually helps to reduce the number of patches required to be reviewed.

\subsection{\rqeffortreduction}

\begin{figure}
\includegraphics[width=0.90\textwidth]{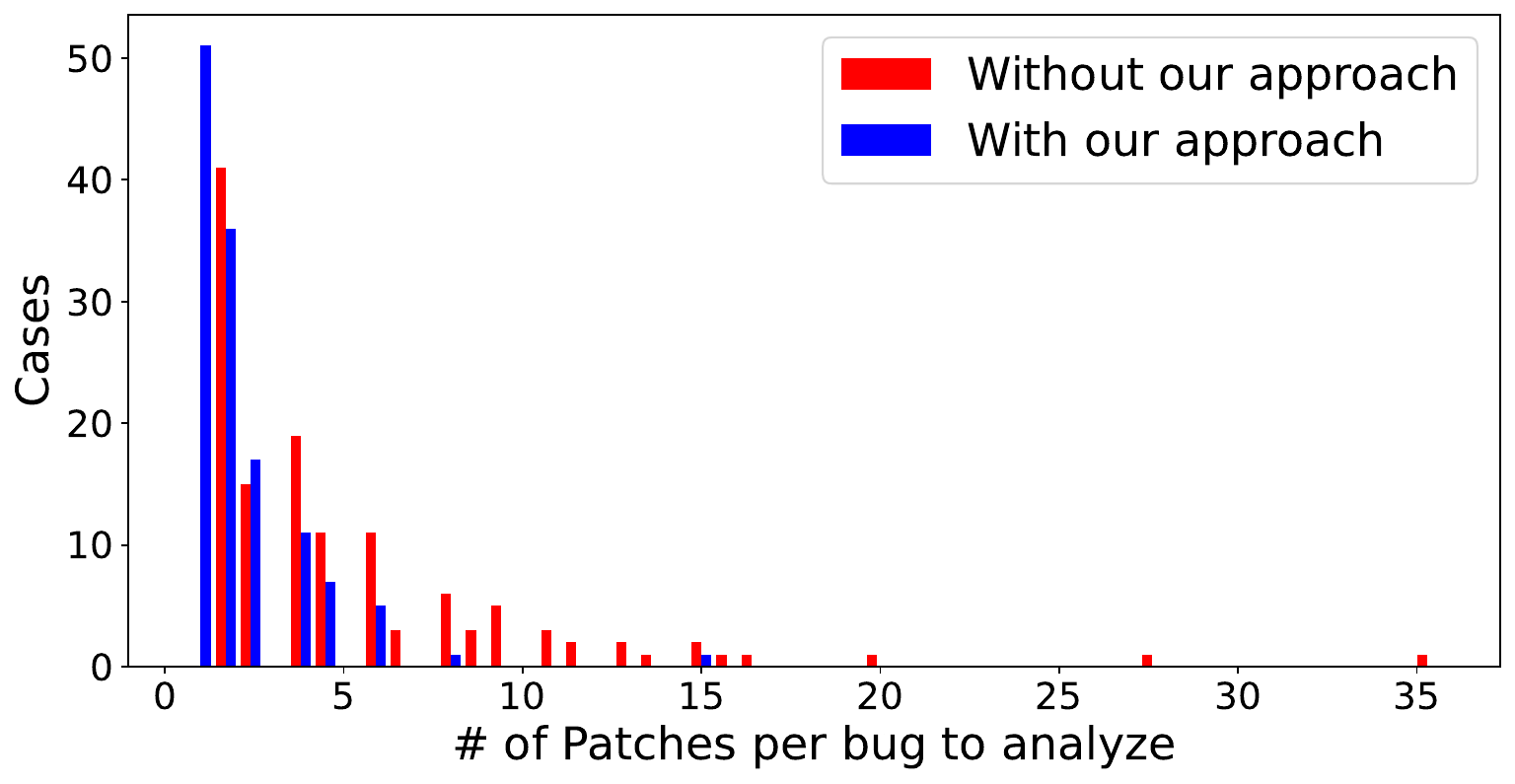}
 \centering
    \caption{RQ2. Distribution of patches to analyze per bug with and without our approach.}
    \label{fig:distPatches}
\end{figure}

We compare the number of patches produced by all selected APR tools vs. the number of patches presented to code reviewer if our approach is used.
Figure~\ref{fig:distPatches} shows the distribution of the number of patches per bug to analyze without and with our approach (red and blue bars, respectively).
We observe that the distribution of patches using our tool (in blue) is distributed more to the left than the other (in red).
More cases on the left mean that the number of patches to analyze is fewer.

Now, we focus on each bug: we study the percentage reduction in the patches to be analyzed when \approach is used vs. reviewing all available patches for a given bug.

The median percentage reduction is 50\% (46.98\%), which means that for half of the bugs \approach reduces the total number of patches one needs to analyze by at least 50\%.

\begin{figure}

\begin{subfigure}[b]{0.5\textwidth}

\includegraphics[width=0.95\textwidth]{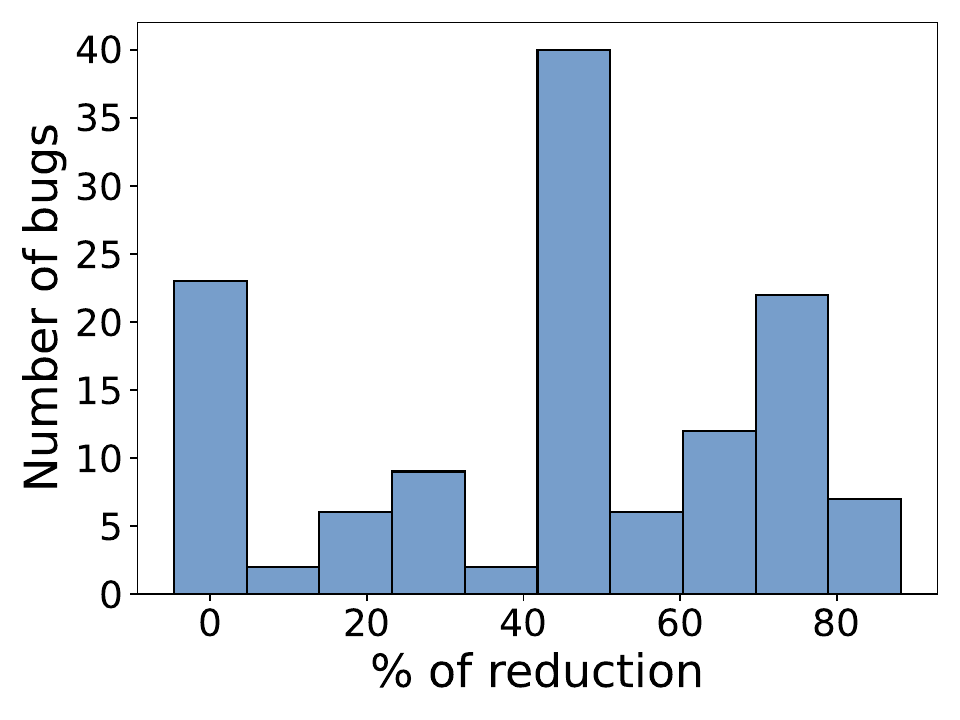}
 \centering
    
\end{subfigure}
\begin{subfigure}[b]{0.5\textwidth}

\includegraphics[width=0.95\textwidth]{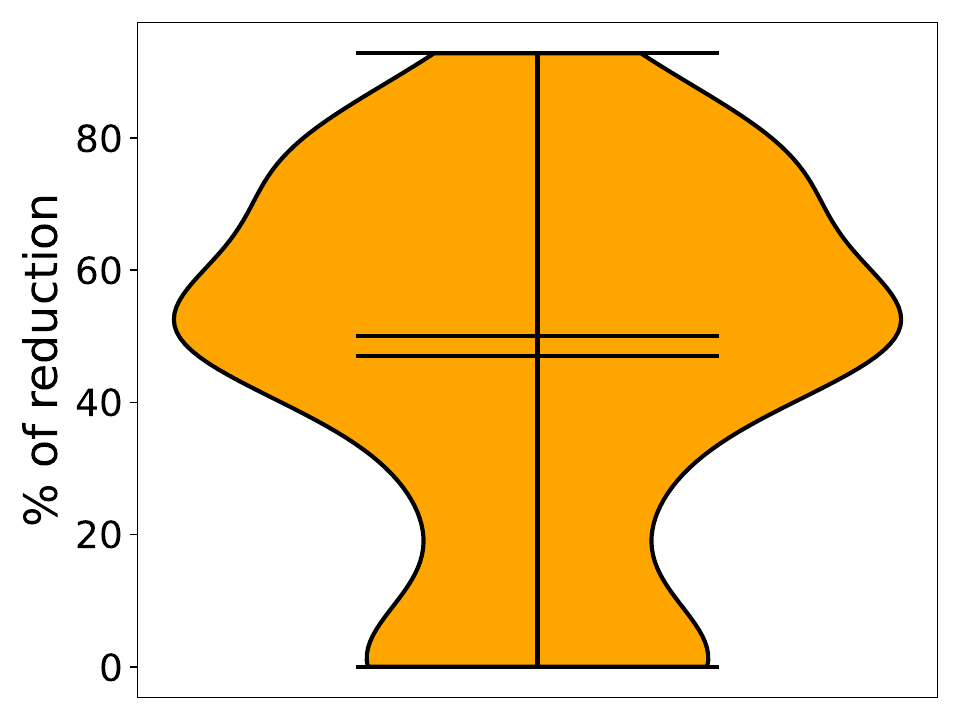}
 \centering

\end{subfigure}

   \caption{RQ2. Distribution of the percentage reduction of the number of patches to review. Reduction for a bug B is computed as follows: ((\#patches for B -  \#clusters for B ) / \#patches for B) $\times$ 100.}
    \label{fig:distReduction}
   \label{fig:distReduction}
\end{figure}

The distribution in Figure~\ref{fig:distReduction} also shows that for 23 bugs we achieve no reduction.
That is, for 23 bugs each cluster contains a single patch, thus all need to be analyzed.
For most of those cases (16), the number of patches (and clusters) is two.
In other words, a \patchreviewer only needs to check two patches per bug.
Even in the cases there is no reduction in terms of number of patches to be analyzed, \approach provides developers, for each of those patches, new inputs (included in the generated test cases) that expose the unique behavior of each of those patches.
Using that information, our approach could help developers decide which patch is better.

\begin{tcolorbox}
{\bf Answer to RQ2.} \approach is able to reduce by a median of 50\% the number of patches to analyze per bug.
Thus \approach could help code reviewers (developers using repair tools or researchers evaluating patches) to reduce the time required for patch assessment.
\end{tcolorbox}

The findings discussed thus far already show that our approach could be very useful to code reviewers.
Firstly, it can significantly reduce the number of patches for review, thus reducing the time and effort required for this task.
Secondly, it can provide code reviewers with test inputs that help differentiate between patches, thus reducing the complexity of patch review.

\paragraph*{{\bf Execution time of \approach.}}

We calculate the execution time of \approach for a given bug by summing the time required for two tasks: 
\begin{inparaenum}[\it 1)] 
\item generating tests for each patch discovered for a given bug (limited to a one-minute time budget), and 
\item executing the newly generated tests for each discovered patch. This results in a total number of executions equivalent to the product of the number of generated tests and the number of discovered patches for a given bug.
\end{inparaenum}

The median execution time per bug is 4.06 minutes (avg. 5.63).
We recall that the process of finding candidate patches is time-consuming \cite{martinez2017automatic,Liu2020OnTE,LIU2021Critical}, taking more than this time to find the first plausible patches, as it requires several compilations and validation using human-written test cases. 
Consequently, the overhead introduced by \approach does not significantly impact the time of the repair process.
Nevertheless, in the Discussion (Section~\ref{discussion}) we discuss potential solutions to further decrease the execution time of \approach.

\subsection{\rqcorrectcluster}
\label{sec:result:correctness}

\begin{table}[]
\centering
\begin{tabular}{lr}
\toprule
{\bf Purity of Clusters} & {\bf \#Bugs} \\
\midrule
Only pure clusters (either correct and incorrect)    & {17}    \\ 
At least 1 mixed cluster      & {21}     \\
Only pure incorrect      (all patches are incorrect)                  & {40}    \\
Only pure correct       (all patches are correct)                      & {0}     \\ 

\midrule
Total (bugs with multiple clusters)   & {78}   \\

\bottomrule
\end{tabular}
\caption{RQ3. Classification of bugs based on cluster purity.
}
\label{tab:purityOfClusters}
\end{table}

\subsubsection{Analysis of bugs with multiple clusters}
\label{sec:correctnessmultcluster}

We now focus on the \bugsmultipleclusters bugs for which our approach produced multiple clusters.
We analyze their purity, i.e., if the patches in a given cluster are all correct or all incorrect.
This will allow us to measure the ability of test-case generation to differentiate between correct and incorrect patches.
Table \ref{tab:purityOfClusters} summarizes our results.
The first two rows group the bugs that have both correct and incorrect patches (17 + 21 = 38 in total).
The third and fourth rows show the number of bugs with either only incorrect patches (40 in total), or only correct patches (zero in total, which means that for every correct patch there is at least one repair tool that generates an incorrect patch).

For 17 out of the 38 bugs (44\%) with both correct and incorrect patches, all clusters generated by \approach are pure.
This means that \approach is able to perfectly distinguish between correct and incorrect patches.
For the remaining 21 bugs (56\%), \approach generates more than one cluster, and at least one of them includes correct and incorrect patches. 
This means that the generated test cases are not capable of detecting the ``wrong'' behavior of the incorrect patches, i.e., they are not capable of detecting, via new inputs, the differences between the behavior of correct and incorrect patches. 
Section~\ref{sec:qualityTest} studies whether this limitation is related to the quality of the newly-generated tests.

Producing only pure clusters has two advantages. When selecting patches from a pure cluster, there is no risk of selecting an incorrect patch over a correct one. 
Moreover, if we know that a given cluster only contains correct patches, we can then safely select a patch from a cluster that satisfies perhaps an additional criterion, such as readability or patch length.

\begin{tcolorbox}
{\bf Answer to RQ3.} 
Based on the generated test cases, \approach is able to cluster plausible patches in a way that a single cluster will contain only correct or incorrect patches for 44\% of the bugs considered (17/38). 
This signifies that for these bugs, it would be sufficient to assess just one of the patches from each cluster in order to check whether it contains a correct patch. Furthermore, a \patchreviewer can choose any patch from a cluster containing correct patches based on additional criteria that best fit their codebase.
\end{tcolorbox}

\subsubsection{Analysis of single cluster bugs}
\label{sec:correctnesssinglecluster}

We previously focused on bugs with multiple clusters and the correctness of the patches that are contained in them.
Now, we focus on bugs for which \approach could not generate more than one cluster.
As Table \ref{tab:clustersummary} shows, there are 51 of such bugs.
For 41 of them (78.4\%), all patches are incorrect. This means that all incorrect patches produce the same behavior in the buggy program according to the test.
The other 11 bugs (21.6\%) contain correct and incorrect patches. 
However, \approach cannot find any behavioral differences using generated test cases, which result is `passing' (i.e., there is no error or failure assertion) on all patches.
We will study such bugs with a single cluster in Section~\ref{sec:qualityTest} to know whether this limitation is related to the quality of the newly-generated tests.

\subsubsection{Relationships between overfitting patches and failing test cases}
\label{sec:correctnesssinglecluster}

We now focus on the behaviors of patches from mixed  clusters on the generated tests.
As we previously mentioned, 21 bugs have one mixed cluster (and one or more pure clusters) and 10 bugs have a single mixed cluster.
On the one hand, from the 21 bugs with multiple clusters, we found  the patch from the mixed clusters fails on one or more newly-generated test cases.
We recall that those failing test cases are provided from patches that belong to a different cluster.
On the other hand, the remaining two bugs have a mixed cluster for which its patches do not fail any newly-generated test cases. 
With respect to the 10 bugs with a single mixed cluster, by construction there is no failing test.

We conclude that in the bugs with multiples clusters, one or them mixed, the behavior of the patches is diverse, and this diversity is \emph{partially} detected by the newly-generated test cases, which are not fully capable of capturing some of the incorrect behaviour from overfitting patches.

\subsection{\rqsota}

We now study to what extent our approach is able to complement two state-of-the-art patch assessment tools: Cache~\cite{Lin2022ContextEmbeeding} and ODS~\cite{ods}.

We first focus on bugs for which \approach generates pure clusters.
We recall those bugs contain one or more clusters with only correct patches and one or more with only incorrect patches, but any cluster with both correct and incorrect.
As discussed in Section \ref{sec:correctnessmultcluster}, those bugs are 17.

After analyzing the assessment done by ODS and Cache on patches from those bugs, we observe that for 15 of those bugs, one of the assessment tools misclassify the correctness of a patch.
In particular, ODS and Cache misclassify patches from 12 bugs and 8 bugs, respectively (three bugs are misclassified by both tools).
For example, for bug  Math-32 from Defects4J, \approach groups patches in two `pure' clusters: one with three incorrect patches, the other with one correct patch. 
ODS incorrectly classifies as overfitting  the correct patch, and marks as correct one of the overfitting patches.
In this case, \approach provides developers `hints' about that misclassification, in particular, let developers know that three patches (two truly incorrect but one wrongly classified as correct) behave similarly according to generated tests (that is, it provides a test case where all the three patches fail one assertion).

\begin{tcolorbox}
{\bf Answer to RQ4.} Our results show that   \approach can complement the overfitting assessment analysis done by state-of-the-art assessment tool: combining patch assessment with cluster classification can expose misclassifications.
\end{tcolorbox}

We also inspect the performance of ODS and Cache on the 21 bugs with multiple clusters and at least one mixed cluster.
For each mixed cluster, we verify whether ODS and Cache are able to successfully recognize the overfitting patches from the correct ones.
We recall that all these patches from a mixed cluster have the same behavior according to the tests generated from \approach.
Interestingly, for 17 of those bugs (81\%), ODS or Cache incorrectly assess at least one of the patches included in a mixed cluster (in particular, ODS does that incorrect assessment for 16 bugs, Cache for 13, and both for 10 bugs).

Finally, we focus on the results of the assessment from  ODS and Cache on the patches from bugs for which \approach creates just a single mixed cluster, i.e., the cluster contains both correct and incorrect patches.
As mentioned in Section~\ref{sec:correctnesssinglecluster} there are ten of such bugs. (The other 41 bugs with a single cluster contain only incorrect patches).

We find that for each of these 10 bugs, ODS or Cache produce at least one incorrect patch assessment: 
ODS fails on at least one patch assessment on nine bugs, and Cache does it on seven bugs.
For example, the dataset from Wang et. al has five plausible patches from Closure-86 bug, all generated by SequenceR.
According to the ground-truth provided by the dataset, only one of them is correct while the other four are incorrect.
Ideally, \approach would have had to generate \emph{pure} clusters, one with the correct patch, and one or more clusters with the incorrect patches.
However, both ODS and Cache could not identify any patch as correct: they wrongly classified all SequenceR patches as overfitting. This classification is a false positive.

ODS and Cache also suffer from false negatives.
For example, there are six plausible patches from bug Math-59, one of them is overfitting (that one generated by SequenceR), while the other five are correct. \approach groups them into a single cluster because it cannot observe behavioral differences between them.
 ODS and Cache cannot observe any differences: they classify all patches as correct. This produces a false negative on the classification of the overfitting patch.

\begin{tcolorbox}
{\bf Answer to RQ4 (cont.)} Our results show that in most of the cases that \approach cannot generate test that difference correct patches from overfitting, two state-of-art patch assessment tools produce  false positive and/or negatives.
This result shows the difficulty of patch assessment and calls for further research.

\end{tcolorbox}


\subsection{\rqqualitytc}
\label{sec:qualityTest}

To respond to RQ5, we first focus on the metrics computed in all test cases, then focus on the relationship between the metrics and the effectiveness of \approach.

\subsubsection{Quality Metrics from test cases}

Table \ref{tab:metricstest} shows the mean, median, and standard deviation of four metrics computed and returned (together with the generated test cases) by Evosuite: 
\begin{inparaenum}[\it 1)]
\item Number of test cases generated for a given patch,  
\item lines of code (LOC) of such test cases, 
\item the line coverage of the generated test cases, and 
\item the mutation score.
\end{inparaenum}
In this section, we focus exclusively on Evosuite for two main reasons: first, as we discuss in {Section~\ref{discussion}} Evosuite outperforms Randoop, secondly, the metrics are automatically generated by Evosuite.\footnote{Collecting data with EvoSuite~\url{https://www.evosuite.org/documentation/tutorial-part-3/}}

\begin{table}[]
\centering
\begin{tabular}{|l|rrr|}
\hline
Metric & Mean& Median & Std Dev\\
\hline
\# Tests per patch &  98.45 & 46 & 114.12 \\
LOC of a generated test cases for a patch&  418.99 & 184 & 961.08 \\
Line coverage &  70\% & 77\% & 20\% \\
Mutation Score & 31\%& 21\%& 26\% \\
\hline
\end{tabular}
\caption{Metrics from the generated test cases for all the patches considered.}
\label{tab:metricstest}
\end{table}

\begin{figure}
     \centering

     \begin{subfigure}[b]{0.42\textwidth}
         \centering
         \includegraphics[width=\textwidth]{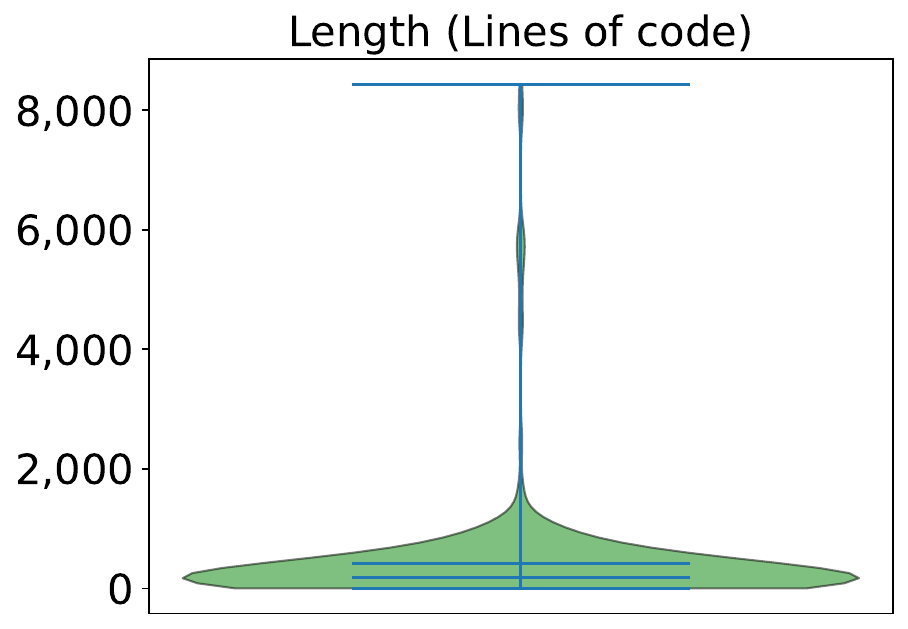}
         \caption{Lines of code from test file}
         \label{fig:violength}
     \end{subfigure}
     \begin{subfigure}[b]{0.42\textwidth}
         \centering
         \includegraphics[width=\textwidth]{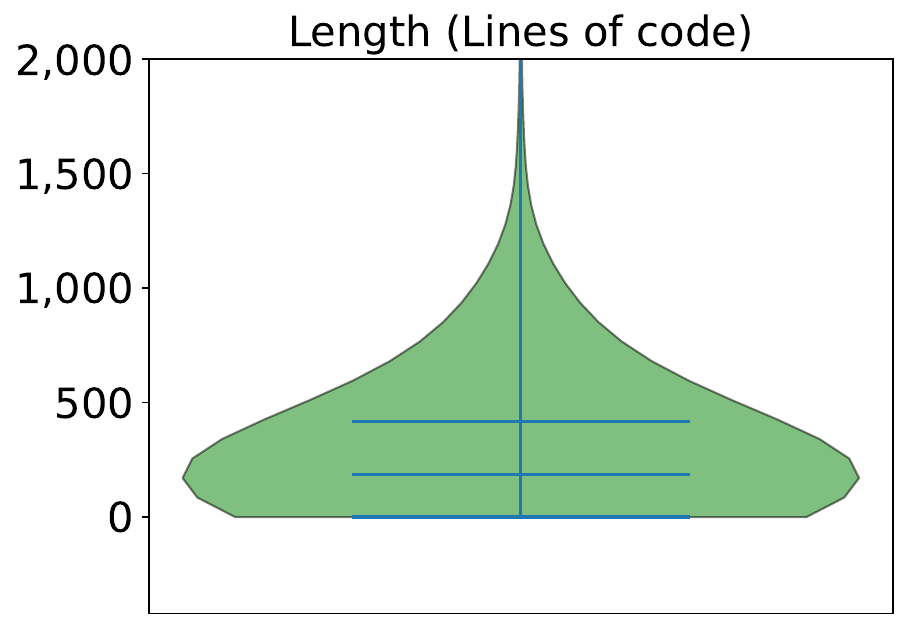}
         \caption{Lines of code from test file (without outliers with length $>$ 2000 LOC)}
         \label{fig:violength_outliers}
     \end{subfigure}

          \begin{subfigure}[b]{0.45\textwidth}
         \centering
         \includegraphics[width=\textwidth]{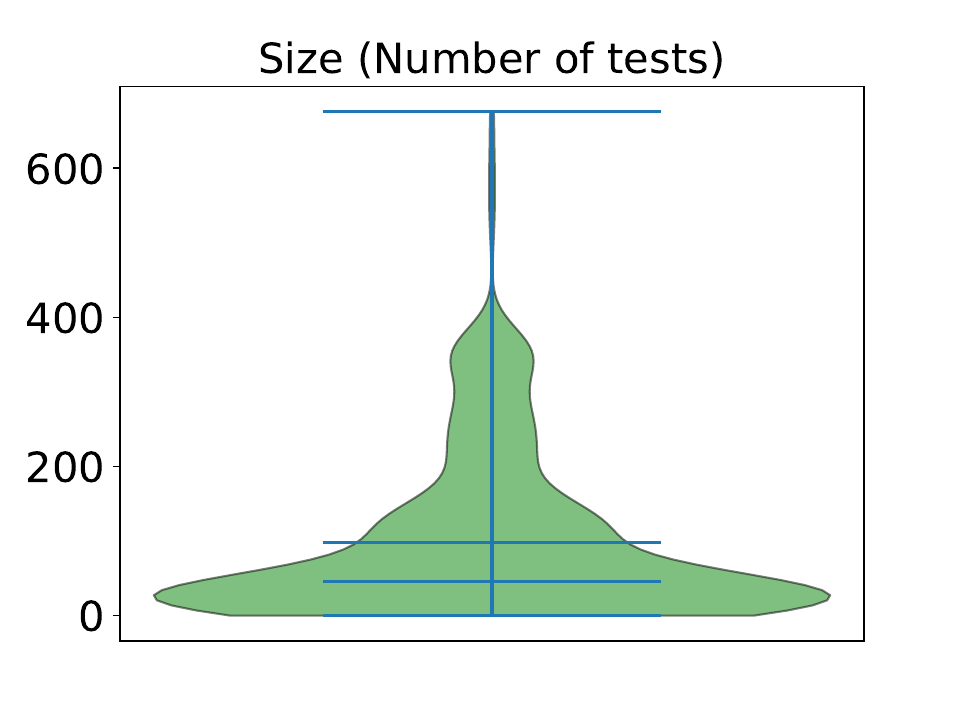}
         \caption{Number of test cases per test file}
         \label{fig:vioSize}
     \end{subfigure}
     \begin{subfigure}[b]{0.45\textwidth}
         \centering
         \includegraphics[width=\textwidth]{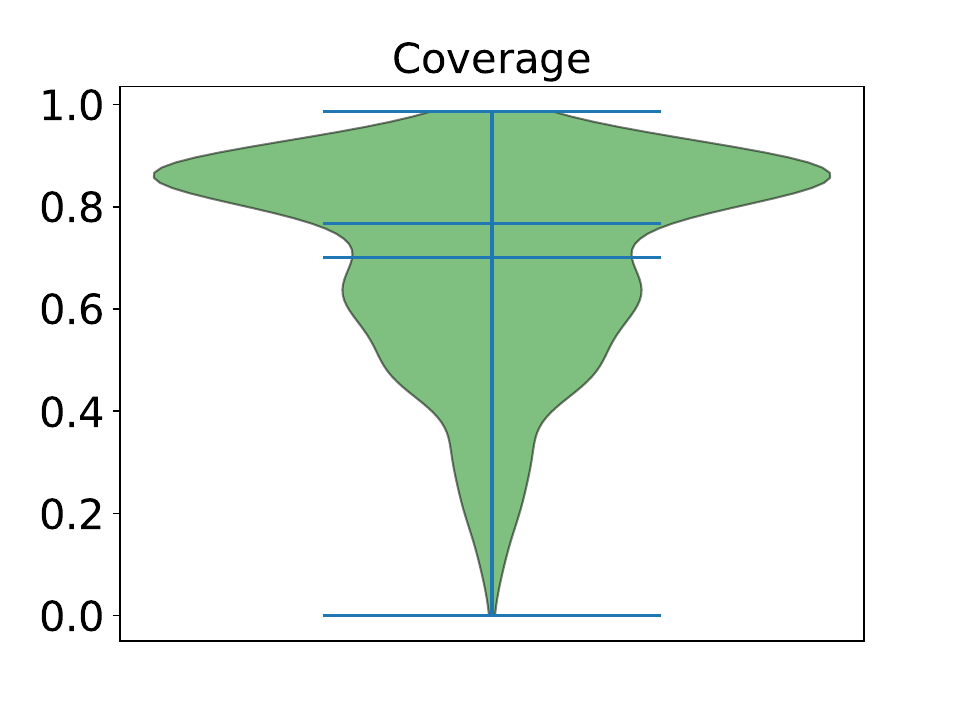}
         \caption{Line coverage}
         \label{fig:viocoverage}
     \end{subfigure} 

     \begin{subfigure}[b]{0.60\textwidth}
         \centering
         \includegraphics[width=\textwidth]{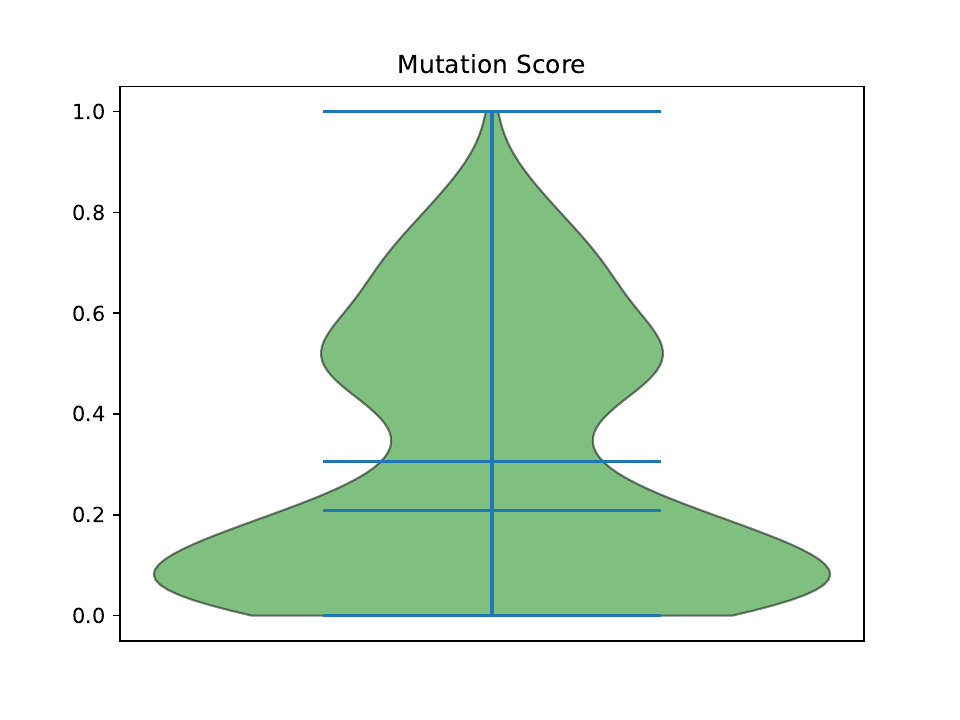}
         \caption{Mutation Score}
         \label{fig:viomutscore}
     \end{subfigure}

   \caption{Metrics computed from newly-generated tests.}
         \label{fig:viotest}  
\end{figure}

\emph{Number of Test cases generated per patch}\footnote{In Evosuite this metrics is reported as ``Size".}:
We observe that the median value of generated test cases is 46 test. 
The average is larger (98.45) because, as we can see in Figure~\ref{fig:vioSize}, there is still a considerable number of patches with between 200 and 400 generated tests.

\emph{Number of lines of code in the test cases}\footnote{In Evosuite this metrics is reported as ``Length".}:
We observe that the median value is 184 lines of code, which includes all the code from test cases generated for a patch. 
The average is also higher (418.45) because, as we can see in Figure~\ref{fig:violength}, for some patches, there are outliers with more than 2000 test cases generated. Figure~\ref{fig:violength_outliers} shows the distribution without those outliers. We observe that most of the test generated have less than 1000 LOC.

\emph{Coverage:}
The coverage reported by Evosuite reaches a median coverage of 77\% but the mean is just lower (70\%).
The distribution presented in Figure \ref{fig:viocoverage} shows that most of the generated tests have a high coverage, between 80\% and 100\%, but still a considerable number of tests reach coverage between 40\% and 0.6\%.

\emph{Mutation Score:}
The mutation score reported by Evosuite reaches a mean of 31\%  and a lower median (21\%). This means that most of the test cases generated are able to kill at most 21\% of the generated mutants, which is notably low.
The distribution presented in Figure \ref{fig:viocoverage} seems to be a bimodal distribution: Most values are concentrated near 10\% and near 55\%. For very few patches, the generated tests reach a mutation score larger than 80\%.

\subsubsection{Relation between Quality metrics and Effectiveness of \approach}

Table \ref{tab:metricscompt} shows the median of each metric, but, as a difference from the previous section, we now consider a subset of tests. 
In particular, in column ``Mixed Clusters" (second column), we consider test cases generated related to bugs with mixed clusters.
In column ``Pure Cluster'' (third column), we consider test cases generated related to bugs with only pure clusters.
The column ``P-value'' (fourth column) shows the p-value returned by the Mann-Whitney U test.
Finally, the last column ``Reject $H_0$'' shows if the Null Hypothesis, which states that there is no significant difference between tests from mixed clusters and tests from pure clusters w.r.t. a metric, is rejected.

We observe that there are significant differences between the median of the number of test, LOC and line coverage, and using the Mann-Whitney U test we reject the null hypothesis for all these three metrics.
For example, the patches assigned to mixed clusters have, as a median, 35 test cases.
However, the patches assigned to pure clusters have, as a median, 126 test cases.

\begin{tcolorbox}
{\bf Answer RQ5:} Our results show that the effectiveness of \approach is related to the quality of the test cases. 
Having a larger number of test cases, larger in terms of LOC and with higher coverage, help \approach to detect behavioral differences between the correct and incorrect patches.
\end{tcolorbox}

\begin{table}[]
\centering
\begin{tabular}{|l| r|r|r|r|}
\hline


{Metric} & {Mixed Cluster}                & {Pure Cluster}              &{P-value} &{Reject $H_0$?}\\ 
\hline
\# Tests per patch & 35.00 & 126.00 & $<$0.0001 & True \\
LOC tests &  142.50 & 392.00 & $<$0.0001 & True \\
Line coverage & 74\%  & 83\% & 0.0039 & True \\
Mutation Score  & 19\% & 28\% & 0.1227 & False \\

\hline
\end{tabular}
\caption{Metrics computed on tests from patches assigned by \approach to Mixed Clusters, and from those assigned to Pure Clusters.}
\label{tab:metricscompt}
\end{table}

\section{Discussion}
\label{discussion}

In this section, we provide a deeper analysis of our findings, presenting two case studies, and an assessment of the performance of the two test generation tools we used in \approach.

\textbf{Case Study: Chart-26 with all pure clusters.}
We collected \newc{11} labelled patches for the bug Chart-26.
After running the generated test cases on these patches, 
\approach created three clusters as follows:
\approach first creates a cluster that has exclusively correct patches. 
Even though the patches are syntactically different, they have the same semantic behaviour. 
For example, a patch from {\sc JAID}~\cite{CPF17} adds an {\tt if}--{\tt return} in \texttt{Axis.java} file, whereas the patches from {\sc TBar}~\cite{LKK19b} add an {\tt if} guard to the same file.
Then, \approach creates two more clusters, both containing only syntactically different incorrect patches.
In one cluster, two patches from {\sc JAID} also affect the \texttt{Axis.java} file but introduce incorrect changes, such as variable assignment. However, in the other cluster, all incorrect patches affect a different file (i.e., \texttt{CategoryPlot.java}).
Overall, \approach not only clustered the correct patches together, but created clusters that contain patches that are semantically similar, despite having syntactic differences.

{\bf Performance of Test-Case Generation Tools.}
\label{sec:discussion:testgeneration}
Our implementation of \approach has a parameter for choosing the test case generation tool to use. The parameter has three values:
\begin{inparaenum}[\it 1)]
\item \emph{EvoSuite}~\cite{evosuite}, 
\item \emph{Randoop}~\cite{randoop},
\item \emph{both} (i.e., it uses the two tools).
\end{inparaenum}
In this paper, we report the results using only Evosuite for different reasons. 
First, \approach using Evosuite achieves better performance than using only Randoop (11 pure clusters).
Second, the use of both tools (i.e., the addition of Randoop to our experimental setup) minimally impacts the performance (just one more bug with pure clusters -Closure 33-) but doubles the execution time.
In conclusion, the addition of new test case generation tools can help \approach to find more pure clusters at the expense of execution time.


{\bf Runtime overhead of \approach.}
We configure test generation tools with a timeout of one minute. 
The execution of the generated test cases takes a few seconds on average. 
Consequently, the overhead mostly depends on the number of generated patches that \approach aims to cluster.

{\bf Reduction of execution time  of \approach.}
The execution time (median 4.06 min per bug) can be reduced by parallelizing the generation of test cases for each patch, rather than sequentially executing them, as we do in this paper (loop in line \ref{algo1:tg:for:file} in Algorithm \ref{algo:testGeneration}).
The execution of these generated test cases can also be parallelized (loop in line \ref{algo1:te:for:testgenerated} in Algorithm \ref{algo:testExecution}).

{\bf Selection of patches from a cluster.} 
All patches grouped in a cluster are semantically equivalent (i.e., behave similarly) according to both the human-written test cases and the new test cases generated by \approach.
Even the approach could randomly pick one of them as the representative of a cluster for the reason mentioned above, it could apply other selection strategies that go beyond the semantics of the patches: i.e., to consider the syntax of patches.
For example, \approach could apply a strategy based on the number of lines of source code it adds and removes from the original code, and favor shorter patches, i.e. those that result in least changed lines in the original code. The preference for shorter patches has also been implemented in previous work (e.g., \cite{Tian2022FailingTestCases}), and has the advantage of helping the understandability and maintainability of the patched code~\cite{Fry2012UnderPatches}.

{\bf Integration with existing APR.}
\approach can be easily integrated with any repair tool that generates Java patches, as it only requires the programs to be repaired and the generated patches as input. In this paper, our tool was used with patches generated by 25 repair tools.

{\bf Parameter fine-tuning.}
In the experiment presented in this paper, \approach invokes the test generation tools (Evosuite and Randoop) using the \emph{default} values of each parameter, including the execution time (aka, \emph{search budget}), which was set to one minute.
Previous work investigated the effect of parameter optimization on test generation using EvoSuite~\cite{ArcuriEvosuiteFinetuning}.
The authors conclude that using
default values is a reasonable and justified choice, whereas parameter tuning (on test generation) is a long and expensive process that might or might not pay off in the end.
Nevertheless, the application of parameter tuning could eventually improve the results presented in this paper.
For example, increasing the execution time beyond the default time budget of 60 seconds provides Evosuite more time to find better quality test cases (in the context of this work, better test quality means, for instance, to strengthen the coverage on lines that have been patched).
Beyond time, there are other parameters that could be fine-tuned with the goal of searching for improving the capacity of \approach, such as the crossover rate, population size, elitism rate, selection, and parent replacement check. The impact of these five parameters on test generation was previously studied by~\cite{ArcuriEvosuiteFinetuning}.

{\bf Limitations.}
\approach uses test generation tools to generate test cases for a patched program version. 
Consequently, the efficacy of \approach heavily depends on the ability of these generation tools to create test cases (that is, a set of inputs and assertions of system output) that exercise buggy behavior affected by a patch.
For example, if a generation tool generates test cases that do not cover a buggy line, then \approach will not be able to differentiate the behavior of the patches and thus will create a single cluster.


\section{Threats to Validity}

Next, we discuss potential issues regarding the implementation of \approach ({\it internal validity}),  the design of our study ({\it construct validity}), and the generalisability of our findings ({\it external validity}).

\textbf{Internal Validity}.
The source code of \approach and the scripts written for generating and processing the results of our experiments may contain bugs.
This issue could have introduced a bias to our results,
by removing or augmenting values.
To mitigate this issue we made our source code, as well as the scripts used for the analysis and processing of the results of our study, publicly available in our repository~\cite{appendix} for external validation.

\textbf{Construct Validity}. There is randomness associated with use of test-case generation tools such as {\sc EvoSuite}~\cite{evosuite} and {\sc Randoop}~\cite{randoop},
because of the nature of the algorithms used in such tools.
Therefore, the results of our experiments could vary between different executions.
In this experiment, we execute {\sc Randoop} and {\sc Evosuite} once on each patch.
Doing more executions of those tools (using different random seeds) could produce more diverse test cases that may find further differences between patches and, consequently, help \approach produce better results.


\textbf{External Validity}. 
Our study uses the dataset from Wang et al., which is in turn composed of two other datasets: DRR~\cite{Ye2021DDR} (automated evaluation, using automated test cases generated on the human-patched program, taking it as ground truth) and the one from Liu et al.~\cite{Liu2020OnTE} (manual evaluation done by humans). These two manners of labeling patches may affect our results.
However, the patches from those work were then re-assessed for  Wang et al, giving as a results the dataset with 902 patches that we use in this experiment.

Durieux et al.~\cite{Durieux:2019:RepairThemAll}
observed that the {\sc Defects4J} benchmark might suffer from overfitting, as it has been used for the evaluation of most {\sc apr} tools available. Thereby it can produce misleading results regarding the capabilities of {\sc apr} tools to generate correct patches.
However, since {\sc Defects4J} has been used for the evaluation of
most {\sc apr} tools, we were able to find labeled data from multiple different {\sc apr} tools only for {\sc Defects4J}.
Further research on the impact of using other bug benchmarks would tackle this threat. 






\section{Related work}
\label{rw}

In this section we discuss the most relevant related work and compare and contrast them to our proposed \approach.


\subsection{Patch clustering}

Similar to \approach, 
Cashin et al. \cite{Cashin2019Undestanding} present PATCHPART, a clustering approach that clusters patches based on invariants generated from the existing test cases. 
Using Daikon~\cite{daikon} to find dynamic invariants and evaluated on 12 bugs (5 from GenProg and 7 from Arja), PATCHPART reduces human effort by reducing the number of semantically distinct patches that must be considered by over 50\%.
Our approach, in contrast, is based on output of the execution of newly generated test cases, and is validated on a larger set of bugs (139 vs 12),  tools (25 vs 2).
A deeper comparison with PATCHPART is not possible, as the information about the bugs repaired, considered patches, clusters generated and the tool is not publicly available. 

Mechtaev et al. \cite{Mechtaev:test-equivalence} provide an approach for clustering patches based on test equivalence.
There are two main differences between their work and ours.
First, the main technical difference is that our clustering approach exploits additional automatically generated test cases, which are created to generate inputs that enforce diverse behavior, while  Mechtaev et al. use solely the existing test suite written by developers, thus is unable to detect differences not exposed by unseen inputs.
Secondly, the goal of our approach also differs: to group patches generated by different (and diverse) repair tools after those are generated, while Mechtaev et al. group patches from a single repair tool as they incorporate their approach to the patch generation process from one tool.
For that reason, our evaluation considers patches from 25 repair tools for Java, while Mechtaev et al. evaluated four repair tools for C.

\subsection{Patch assessment}


Wang~et~al.~\cite{wang2020AutomatedAssessment} provide an overview and an empirical comparison of patch assessment approaches.
One of the core findings of Wang~et~al.~\cite{wang2020AutomatedAssessment} is that existing techniques are highly complementary to each other.
These overfitting techniques can be used to complement our work.
For instance, we can apply \approach, which is based on the cross-execution of newly generated test cases, on a set of previously filtered patches using automated patch assessment, or to use a patch ranking technique on the clusters created by \approach.
We describe a selection of such approaches in this section.
Here we divide work on automated patch assessment into two categories: approaches that focus on overfitting as: (1) {\it independent tools}, which can be used with different {\sc apr} approaches; (2) {\it dependent tools}, which are incorporated into specific repair tools.

{\bf Independent Approaches.
} 
{\sc Opad} is a dynamic approach, which filters out overfitted patches by generating new test cases using fuzzing.
Initially, {\sc Opad} was developed for C programs and evaluated on {\sc GenProg}, {\sc Kali}, and {\sc SPR}~\cite{Opad}. Recently, a Java version for {\sc Opad} has been introduced by Wang et al.
~\cite{wang2020AutomatedAssessment}.
There are several tools that use patch similarity for patch overfitting assessment.
For instance, {\sc Patch-sim}~\cite{patchsim} and {\sc Test-sim}~\cite{patchsim} have been developed for Java and evaluated on {\sc jGenProg}, {\sc Nopol}, {\sc jKali}, {\sc ACS} and {\sc HDRepair}.
Specifically, the approach generates new test inputs to enhance original test suites, and uses test execution trace and output similarity to determine patch correctness.
{\sc ObjSim}~\cite{Ghanbari2020ObjSim} employs a similar strategy and has been evaluated on patches generated by {\sc PraPR}.
The above approaches involve code instrumentation, which can be costly.
Like many other approaches, they also provide patch ranking as an output. 
{\sc DiffTGen} \cite{xin2017identifying} identifies overfitted patches by generating new test inputs that uncover semantic differences between an original faulty program and a patched program.
Their test cases are created from an oracle, and in the evaluation of DiffTGen, the authors use the human-written patches as correctness oracle.
Unlike them, in our work we do not assume existence of an oracle because it is not available during the repair process.
Our approach creates new test cases from generated patches (not from human-written patches) and analyze the behavioural differences between them (not between a candidate patch and a human-written patch).
The goals of our technique \approach and the mentioned techniques are different. 
Those aim to classify patches as overfitting (in order to remove them), our technique clusters patches according to their behavior. Consequently, even if both aim to reduce human effort, the final goals are different.

Other patch assessment approaches that use machine learning (ML) to label correct and incorrect patches have recently emerged.
For instance, {\sc ODS} is a novel patch overfitting assessment system for Java that leverages static analysis to compare a patched program and a buggy program, and a probabilistic model to classify patches as correct or not~\cite{ods}.
{\sc ODS} can be employed as a post-processing procedure to classify the patches generated by different {\sc apr} systems.
Tian~et~al.~\cite{tian2020evaluating} demonstrated the potential of embeddings to empower learning algorithms in reasoning about patch correctness: a machine learning predictor with the {\sc BERT} transformer-based embeddings associated with logistic regression.
Tian~et~al. also propose BATS~\cite{Tian2022FailingTestCases}, an unsupervised learning-based approach to predict patch correctness by statically checking the similarity of generated patches against past correct patches.
In addition to the patch similarity, BATS also considers the similarity between the failing test cases that expose the bugs fixed by generated patches and those failing test cases that expose the bugs repaired by past correct patches.
Kang et al. \cite{kang2022language}  propose an approach based on language models which prioritizes patches that generate natural code.
Cache from Lin et al.~\cite{Lin2022ContextEmbeeding} presents a deep learning-based classifier to predict the correctness of the patch.
The evaluation of Cache shows that it performs better than the previous mentioned approaches including ODS and {\sc Patch-sim}~\cite{patchsim}.
Liang et al.~\cite{Liang2021} present an interactive filtering approach to patch review, which filters out incorrect patches by asking questions to the developers. The authors implemented this approach in an Eclipse plugin tool called InPaFer.
The main differences between these machine learning approaches and \approach, are as follows: (1) we aim to cluster patches based on behavioural differences, while the aforementioned approaches try to detect overfitting patches (2) the ML-based approaches are static (do not execute the patches), while our approach performs dynamic analysis by executing the generated patches using newly generated test cases (3) we do not require existence of a dataset of previously fixed patches.

{\bf Dependent Approaches.
} 
This category includes {\sc apr} tools that also implement patch overfitting assessment techniques.
Most techniques are based on static analysis and relevant heuristics. For instance, {\sc S3} is an {\sc apr} tool for the C programming language that uses syntax constraints for assessing patch overfitting~\cite{s3}.
{\sc ssFix} is an {\sc apr} tool for Java that leverages syntax constraints for assessing patch overfitting~\cite{ssFix}.
{\sc CapGen} considers programs' {\sc ast}s and a context-aware approach for assessing patch overfitting~\cite{wen2018Contextaware}.
{\sc Prophet} is an {\sc apr} tool for C programs   that rank patch candidates in the order of likely correctness using a model trained from human-written patches.
Other techniques apply dynamic strategies for filtering overfitting patches.
For instance, {\sc Fix2Fit}~\cite{Gao:crashprepair} is an {\sc apr} tool for C programs that defines a fuzzing strategy that filters out patches that make the program crash under newly generated tests.
Our approach can complement these tools: \approach can receive as input the (filtered) patches from one or more of those {\sc apr} tools, and present to the user only those that behave differently.

\section{Conclusions}

We have introduced \approach that is able to reduce the amount of patches required to be reviewed.
\approach clusters semantically similar patches together by exclusively utilising automated test generation tools.
In this paper, we evaluate it in the context of automated program repair. 
APR tools can generate multiple plausible patches, that are not necessarily correct.
Moreover, different tools can fix different bugs.
Therefore, we consider a scenario where multiple tools are used to generate plausible patches for later patch assessment.

We use 902 patches from previous work that were labeled as correct or not and evaluated \approach using that set.
We show that \approach can indeed cluster syntactically different yet semantically similar patches together.
  
Results show that \approach reduces the median number of patches that need to be assessed per bug by half. 
This can save significant amount of time for developers that have to review the multitude of patches generated by {\sc apr} techniques.
Moreover, we provide test cases that show the differences in behavior between different patches.

For 44\% of the bugs considered, all clusters are pure, i.e., contain only correct or incorrect patches. 
Thus, \patchreviewer can select any patch from such a cluster to establish correctness of all patches in that cluster.

In future work we will study the feasibility of complementing our approach with other techniques, e.g., patch ranking, in order to help \approach to select the most adequate patch from a given cluster.


\section*{Fundings}
We would like to gratefully acknowledge support from: Ramony Cajal
Fellowship no. RYC2021-031523-I, ERC Advanced Grant no. 741278, UKRI EPSRC grant no. EP/P023991/1, Australian Research Council grant no. DP210100041. 

\section*{Conflict of Interests}
The authors declare that they have no conflict of interests.

\section*{Copyright}
To satisfy UKRI funder requirements, for the purpose of open access, the authors have applied a Creative Commons Attribution (CC BY) license to any accepted manuscript version arising.

\bibliographystyle{IEEEtran}
\bibliography{refs}

\end{document}